\documentclass[prd,showpacs,nofootinbib,preprintnumbers]{revtex4}

\usepackage{latexsym}
\usepackage{amssymb}
\usepackage{amsmath}
\usepackage{bm}
\usepackage{subfigure}
\usepackage{float}
\usepackage{indentfirst}
\usepackage{array}
\usepackage{xspace}
\usepackage[dvips]{epsfig}
\usepackage{psfrag}
\usepackage{amscd}
\usepackage{slashbox}
\usepackage{hhline}

\newcommand{\comment}[1]{}
\newcommand{\vm}[1]{\vphantom{#1}}
\newcommand{\hsp}{\hspace{0.15em}}
\newcommand{\nhsp}{\hspace{-0.15em}}
\newcommand{\smhsp}{\hspace{0.06em}}

\newcommand{\ds}{\displaystyle}

\newcommand{\lbr}{\left(}
\newcommand{\rbr}{\right)}

\newcommand{\la}{\lambda}

\newcommand{\vak}{\varkappa\,}
\newcommand{\pa}{{\partial}}
\newcommand{\ga}{\gamma}
\newcommand{\si}{\sigma}
\newcommand{\Si}{\Sigma}

\newcommand{\fr}{\frac}

\newcommand{\bw}{\begin{widetext}}
\newcommand{\ew}{\end{widetext}}
\newcommand{\be}{\begin{align}}
\newcommand{\ee}{\end{align}}
\newcommand{\ba}{\begin{eqnarray}}
\newcommand{\ea}{\end{eqnarray}}

\newcommand{\h}{H}

\def\cV{{\cal V}}

\def\e{{\rm e}}

\newcommand{\zt}{\dot{z}}

\newcommand{\sgn}{{\rm sgn\smhsp}}
\newcommand{\nn}{\nonumber}

\newcommand{\br}{\mathrm{br}}

\newcommand{\ah}{ \mathrm{a}}
\newcommand{\bh}{ \mathrm{b}}

\newcommand{\smph}{\vphantom{ d^0_0}}
\newcommand{\vp}{\vphantom{\frac{a}{a}}}

 \numberwithin{equation}{section}

\begin{document}

\title{Energy-momentum balance in a particle\,{\kern 1pt}domain wall perforating collision}
\author{D.\,V.\,Gal'tsov$^{1}$,
 E.\,Yu.\,Melkumova$^1$ and
P.\,Spirin$^{1,2}$\footnote{galtsov@phys.msu.ru,
elenamelk@physics.msu.ru, pspirin@physics.uoc.gr.}} \affiliation{
${}^1$\;Department of Theoretical Physics, Faculty of Physics,
Moscow State University,  119899, Moscow, Russia; \\
${}^2$\;Institute of Theoretical and Computational Physics,
Department of Physics, University of Crete, 71003, Heraklion,
Greece. }


%\date{\today}

\begin{abstract}
We investigate the energy-momentum balance in the perforating
 collision of a point particle with an infinitely thin planar domain
wall within the linearized gravity in arbitrary dimensions. Since
the metric of the wall increases with distance, the wall and the
particle are never free, and their energy-momentum balance
involves not only the instantaneous kinetic momenta, but also the
nonlocal contribution of gravitational stresses. However, careful
analysis shows that the stresses can be unambiguously divided
between the colliding objects leading to definition of the
gravitationally dressed momenta. These take gravity into account
in the same way as the potential energy does in the
nonrelativistic theory, but our treatment is fully relativistic.
Another unusual feature of our problem is the nonvanishing flux of
the total energy-momentum tensor through the lateral surface of
the world tube. In this case the zero divergence of the
energy-momentum tensor does not imply conservation of the total
momentum defined as the integral over the spacelike section of the
tube. But one can still define the conservation law
infinitesimally, passing to time derivatives of the momenta. Using
this definition we establish the momentum balance in terms of the
dressed particle and wall momenta.

\end{abstract}
\maketitle

%\tableofcontents

\section{Introduction}

In the standard theory of particle collisions, both  classical and
quantum, one assumes the existence of asymptotic states in which the
particles can be regarded as noninteracting. This gives rise to the
energy-momentum conservation playing a crucial role in the
understanding of such processes.  For this picture to be  valid, the
interaction force between the colliding objects must fall down
sufficiently fast with the distance. Meanwhile, in various
physically interesting situations this is not so, the notable
example being interaction between quarks.

To explore the possibility of the energy-momentum definition in the
absence of asymptotically free states we consider here a collision
of the gravitationally interacting infinitely thin domain wall and
point particle.  Such a problem is of interest for applications in
the standard
\cite{linear,linear1,Vilsh,ChamEard,Stojkovic:2005zh,Flachi:2007ev}
and the Rundall-Sundrum-type \cite{RS1,RS2,Tanahashi:2011xx,BBHS3}
cosmology, string theory \cite{cvet}, in studying brane -black hole
composites \cite{BBHS4,BBHS5}, black hole escape from branes
\cite{escape,escape1,escape2}, and in other situations.
Gravitational force exerted upon the particle by the plane domain
wall does not fall with distance \cite{IS,IS1}, so the particle
cannot be considered free at any moment. If the domain wall is
viewed as a fixed source of gravity, the particle moves along the
geodesic in the space-time generated by the domain wall, and the
notion of the gravitational potential energy can be introduced. But
if one wants to treat both objects on equal footing, the interaction
potential cannot be introduced.

The domain wall  -particle scattering problem, however, is well
posed within the linearized gravity, where it can be formulated  in
close parallel to the case of two gravitating particles. Moreover,
while the   head-on collision of particles is a singular problem
even in the linearized gravity, our process is still tractable,
since the gravitational force acting upon the particle remains
finite when it comes unto contact with the wall. Recently we have
shown \cite{GMZ,GaMeS1} that the  perforation of the domain wall by
the particle can be well described in  linearized gravity in terms
of distributions. A novel feature of this situation is due to the
existence of the internal dynamics of the domain wall which gets
excited after the perforation in the form of the spherical branon
wave.

Here we would like to show that the problem of the energy-momentum
conservation in the domain wall -particle interaction is also
tractable going beyond the linear theory up to the second order in
the gravitational constant. This is needed in order to introduce the
effective gravitational stress tensor which has to be taken into
account in establishing the energy-momentum balance. Such a stress
tensor obtained by expanding the Einstein tensor up to the second
order in metric deviations  is  nonlocal. But, as we will show,
careful analysis allows one to unambiguously split it between the
domain wall and the particle leading to a definition of
gravitationally dressed colliding objects. This dressing resembles
introduction of the potential energy in the nonrelativistic theory,
but an essential difference is that now the treatment is fully
relativistic and both objects are considered on equal footing.
Gravitational dressing does not mean taking into account a proper
gravitational field of each object, but rather accounting for the
gravitational field of the partner. Therefore, our dressing must not
be confused with the self-energy problem.

\section{The Setup}\label{sect2}
Our system consists of a point particle moving along the world line
$x^M=z^M(\tau)$ and an infinitely thin domain wall filling the world
volume $\cV_{D-1}$  given by the embedding equations
$x^M=X^M(\sigma^\mu)$  in $D$ -dimensional space-time with the
metric $g_{MN}$, $M=0,...,D-1,\;\mu=0,...D-2$ of the signature
$(+,-,...,-)$. The action can be written as
\begin{align}\label{Ac}
S=S_{\rm p}+S_{\rm dw}+ S_{\rm grav}\,,
\end{align}
where $S_{\rm p}(z^M,\,e)$ is the particle action in the Polyakov
form
\begin{align}\label{Acp}
S_{\rm p}=- \frac{1}{2} \int \! \left(e\; g_{MN}\dot{z}^M
\dot{z}^N+\frac{m^2}{e}\right) d\tau\,,
\end{align}
[$e(\tau)$ is the einbein on  the particle world line], $S_{\rm
dw}(X^M,\,\gamma_{\mu\nu})$ is the domain wall geometrical action
\begin{align}\label{Acd}
S_{\rm dw}=-\fr{\mu}{2}\int\left[\vp
  X_\mu^M X_\nu^N g_{MN}\gamma^{\mu\nu}-(D-3)\right]\sqrt{|\gamma|}\:d^{D-1}
  \sigma \,,
\end{align}
where $X_\mu^M=\pa X^M/\pa\hsp\sigma^\mu$ are the tangent vectors
and $\gamma^{\mu\nu}$ is the inverse metric on the domain wall world
volume $\cV_{D-1}$, $\gamma={\rm det} \gamma_{\mu\nu}$, and $S_{\rm
grav}(g_{M})$ is the Einstein-Hilbert action reads
\begin{align}\label{Acg}
S_{\rm grav} =  -\frac{1}{\vak^2}\!\int\! R_D \,\sqrt{|g|}\; d^D x
\,,\qquad \vak^2\equiv 16\pi G_D\,.
 \end{align}

Variation of (\ref{Acd}) with respect to $X^M$ and $\gamma^{\mu\nu}$
gives the brane equation of motion in the covariant form
 \begin{align} \label{em}
\partial_\mu\left(  X_{\nu}^N
g_{MN}\gamma^{\mu\nu}\sqrt{|\gamma|}\right)=\frac{1}{2}\,
g_{NP,M}X^N_{\mu} X^P_{\nu}\gamma^{\mu\nu}\sqrt{|\gamma|}\,,
 \end{align}
and the constraint equation
 \begin{align} \label{con} \lbr X_\mu^M X_\nu^N - \fr12\,
\gamma_{\mu\nu}\gamma^{\la\tau} X_\la^M X_\tau^N \rbr g_{MN}
+\fr{D-3}{2}\,\gamma_{\mu\nu}=0\,,
 \end{align}
 whose
solution defines $\gamma_{\mu\nu}$ as the induced metric on
$\cV_{D-1}$: $$ \gamma_{\mu\nu}=X_\mu^M  X_\nu^N g_{MN}{\big
 |}_{x=X}\,.
$$

Similarly, variation of (\ref{Acp}) with respect to $z^M(\tau)$
and
 $e(\tau)$  gives the geodesic equation in
arbitrary parametrization
\begin{align}\label{eomp}
\fr{d}{d\tau}\lbr e \dot{z}^N g_{MN} \rbr=\fr{e}2 \; g_{NP,M}\smhsp
\dot{z}^N \dot{z}^P ,
\end{align}
 and the constraint
\begin{align}
\label{consp} e^2  g_{MN} \dot{z}^M \dot{z}^N=m^2\,.
\end{align}
We prefer to  keep the Lagrange multipliers explicitly to facilitate
formulation of the perturbation theory.

Finally, variation of the Einstein-Hilbert action (\ref{Acg}) over
$g_{MN}$ leads to the Einstein equations
\begin{align}\label{Eeq}
G^{MN}=\frac{1}{2}\,\vak^2 \left[\vp T^{MN}+ \bar T^{MN} \right]\,,
\end{align}
 with the source terms due to the domain wall
\begin{align}\label{EMTw}
T^{MN}=\mu\int X^M_\mu X^N_\nu
\gamma^{\mu\nu}\;\frac{\delta^D\left(x-X (\si)\smph
\right)}{\sqrt{|g|}}\;\sqrt{|\gamma|}\;d^{D-1}\si\, ,
 \end{align}
and the particle (the corresponding quantities here and below will
be labeled by bar):
 \begin{align}\label{EMTp}
 \bar T^{MN}= e \int \fr{\zt^M \zt^N \delta^D\!\left(x-z(\tau)\smph
\right)}{\sqrt{|g|}}\,d\tau \,.
 \end{align}

Einstein equations with the source term (\ref{EMTw}) have some exact
nonsingular solutions  \cite{time,Dokh,IS,IS1,linet}, while no such
solutions are possible for the point particle source. Actually,
reasonable exact solutions exist for branes embedded into
 space-time with codimensions one and two, but not higher. In
any case we need here time-dependent solutions describing the
collision, which can only be constructed perturbatively. We work in
 linearized gravity assuming smallness of deviation of the
space-time metric from  Minkowskian:
\begin{align}\label{meka}
g_{MN}=\eta_{MN}+ \vak \h_{MN}\,,
\end{align}
but we keep  the full Einstein action to be able to extract the
gravitational stress tensor as the second-order expansion term of
the Einstein tensor:
\begin{align}
\label{SMN} G^{MN}=-\fr{\vak}{2}\, \Box
\!\left(\h^{MN}-\frac{1}{2} \,\eta^{MN}\h\right) -\fr{\vak^2}{2}\,
{\sf{S}}^{MN}+O(\h^3)\,,
\end{align}
where $\h=\h^M_M$, $\Box =\eta^{MN}\partial_M\partial_N$, and
${\sf{S}}^{MN}$ stands for the quadratic terms in $\h_{MN}$:
\begin{align}
\label{natag_0} &{\sf{S}}^{MN}  =   2 \hsp  {\h}^{MP , Q}
\h^{N}{}_{\![Q , P]} + {\h}_{PQ}\left(\h^{MP ,NQ}+ \h^{NP , MQ}-
\h^{PQ, MN}- \h^{MN , PQ}\vp \right) -2\hsp \h^{(M}_{P}
\Box \h^{N)P}_{\phantom{P}} -  \nn\\
  & -\frac{1}{2}\,  {\h}^{PQ  , M} \h_{PQ}{}^{,
N}+\frac{1}{2}\, {\h}^{MN}\Box \h + \frac{1}{2}\,\eta^{MN}\!\left(2
{\h}^{PQ}\Box \h_{PQ}- {\h}_{PQ , L} \h^{PL , Q}+\frac{3}{2}\,
{\h}_{PQ , L} \h^{PQ,L}\right).
\end{align}
In this definition there is the following subtlety. The metric
deviation $\h_{MN}$ is defined initially as generally covariant
quantity with {\em lower} indices and then identified with the
Minkowskian tensor whose indices are raised with the inverse
Minkowski metric. The quadratic tensor ${\sf{S}}^{MN}$, which is
also further regarded as the Minkowskian tensor, is obtained
expanding the Einstein tensor with {\em upper} indices, all internal
contractions of metric deviations being performed with Minkowski
metric.

 The full set of
variables in our problem consists of
$z^M(\tau)\;,e(\tau)\;,X^{\mu}(\sigma)\;,\gamma_{\mu\nu}$, and
$\h_{MN}(x)$. To treat the  problem perturbatively we expand all of
them in powers of $\vak$  and derive the system of iterative
equations. The $D$-dimensional Cartesian coordinates of the
embedding space-time are split as $x^M=(x^{\mu},z)$,
$x^{\mu}=(t,\bf{r})$, and the particle is assumed to move along $z$,
i.e. normally to the domain wall. In the zeroth order the particle
is assumed to move with the constant velocity
$$u^M=\gamma(1, 0, ..., 0, v)\,,\qquad {\rm where}\quad
\gamma=1/\sqrt{1-v^2}\,,$$ so the world line and the einbein are
$$  z^{M}(\tau)= u^M\tau \,,\qquad
e={\rm const}=m\;,
$$ corresponding to the parametrization in terms of the proper
time. The wall in the zeroth order is assumed to be plane, unexcited
and being at rest at $z=0$ in the chosen Lorentz frame:
$$ X^M =\Si^M_\mu \si^\mu\,, $$
 where $\Si^M_\mu$ are $(D-1)$ constant Minkowski vectors
 normalized as
 \begin{align}
\Si^M_\mu\Si^N_\nu\eta_{MN}=\eta_{\mu\nu}\,.
 \end{align}
 Obviously, this is a solution to the Eq. (\ref{em}) for $\vak=0$,
and the corresponding induced metric is the four-dimensional
Minkowski metric $\gamma_{\mu\nu}=\eta_{\mu\nu}$. Thus it is
convenient to fix $\Si^M_\mu=\delta^M_\mu$ without loss of
generality. The moment of perforation of the wall by the particle
that occurs at $z=0$ is $t=\tau=0$.

 The metric deviation must be further expanded in $\vak$:
\begin{align} \h^{MN}=h^{MN}+{\bar h}^{MN}+\delta  \h^{MN}\,,
 \end{align} where the
first-order term is split into the sum of contributions of the wall
$h^{MN}$ and of the particle ${\bar h}^{MN}$. These obey the linear
equations
 \begin{align}
 \Box h^{MN} = -\vak
\left(T^{MN}-\frac{1}{D-2}\,\eta^{MN}T_P^P\right)\,,\qquad \Box
{\bar h}^{MN} = -\vak \left({\bar T}^{MN}-\frac{1}{D-2}
\,\eta^{MN}{\bar T}_P^P\right)\,,
 \end{align}
  where the sources must be constructed
in terms of the above zeroth order quantities and the
Fock-de\,Donder gauge $\partial_N h^{MN}-1/2{\kern 1pt}
\partial^M h=0$  is assumed. The next order metric deviation $\delta \h^{MN}$
 does not split  anymore on separate contributions and obeys (in
the same gauge) the d'Alembert equation
\begin{align}
  \Box\left( \delta\h^{MN}-\frac{1}{2}\,\eta^{MN}\delta\h\right) = -\vak \tau^{MN}\,,
  \label{psi2eq}
\end{align}
 with the source  \begin{align} \tau^{MN}=\delta T^{MN}+ \delta{\bar T}^{MN}+S^{MN}(h,\bar
h)\,,\label{source22}
 \end{align}
where $\delta T^{MN}\,, \delta{\bar T}^{MN}$ are the perturbations
of the wall and particle stress tensors, while $S^{MN}(h,\bar h)$
stands for the quadratic form ${\sf{S}}^{MN} $ in which $\h^{MN}$
must be taken as the sum $\h^{MN}=h^{MN}+{\bar h}^{MN}$ keeping only
the crossed terms in $h^{MN}\,,{\bar h}^{MN}$.
  The quantity
$S^{MN}(h,\bar h)$ is regarded as the gravitational stress tensor
whose presence is needed to ensure the fulfillment of the
conservation equation up to the first order in $\vak$:
 \begin{align} \label{contau}
\partial_N \tau^{MN} =0\,.
 \end{align}
Generically the gravitational stress tensor is  a nonlocal quantity,
but, as we will show below, within the perturbation theory it  can
still be split into two contributions which may be attributed to the
wall and the particle separately. This is how the idea of
gravitational dressing is implemented.

The domain of validity of our perturbation theory is somewhat subtle
and worthwhile being discussed in detail.  Gravity force exerted by
the wall upon the particle is repulsive and we consider  the case
when the initial velocity of the particle is large enough to reach
the wall and to perforate it. After the perforation the particle
gets accelerated by the wall's gravitational repulsion and goes
away. Since the metric deviations caused by the wall  in the
linearized gravity are growing with $z$, one can treat the collision
perturbatively only in some vicinity of the perforation moment. From
the particle energy $\mathcal{E}=m\gamma$, the wall's tension $\mu$
(of dimensionality length$^{-(D-1)}$) and the gravitational coupling
constant $\vak^2$, having in $D$ dimensions the dimensionality
length$^{D-2}$, one can form two length parameters (in the units
$c=1$):
\begin{align}
l \simeq  [\vak^2 \mu]^{-1}\,,\qquad r_S \simeq \left( \vak^2
\mathcal{E} \right)^{\frac1{D-3}}\,,
 \end{align}
the first of which corresponds to the curvature radius of the bulk
generated by the wall, while  the second is the gravitational radius
of the energy $\mathcal{E} $. The wall's gravity is small at the
distances from the wall  $z< l\,,$  while gravity of the particle is
small for
  $z^2+{r}^2> r_S^2$.    If
  $r_S\ll l$,   these conditions intersect
within some matching zone. But it turns out \cite{GaMeS1} that,
assuming the linearized gravity to be true gravity theory for point
particles {\em elsewhere} one is still able to treat the collision
up to $z=0$ (the perforation point) consistently in terms of
distributions (i.e. in the formal limit $r_S=0$. Here we will show
that this treatment is consistent with the energy-momentum
conservation in the perforation process in the linear order in
$\vak$, thus giving further evidence for validity of our approach.
Note that this is quite different from the more singular case of the
head-on collision of two gravitating point particles that cannot be
treated within the linearized gravity.

\section{First-order perturbations}
For reader's convenience we briefly reproduce here the results
obtained in \cite{GaMeS1}. The full metric deviation in the first
order is the sum of $ \bar{h}_{MN}$ generated by the unperturbed
particle motion and $ h_{MN}$ representing gravity of the
unperturbed wall at rest. The first reads explicitly for $D>3$:
\begin{align} \label{hpart}
 \bar{h}_{MN}(x)=-\fr{\vak
\,m\Gamma\left(\fr{D-3}{2}\right)}{4\pi^{\fr{D-1}{2}}}
 \left(u_M
u_N-\fr{1}{D-2}\,\eta_{MN}\right)\fr{1}{[\gamma^2(z-v
t)^2+r^2]^{\fr{D-3}{2}}}\,,
 \end{align}
where $r= \sqrt{\delta_{ij} \sigma^i \sigma^j}$ is the radial
distance on the wall from the perforation point. This is just the
Lorentz-contracted $D$-dimensional Newton  field of the  uniformly
moving particle. In what follows we will also need the corresponding
Fourier transform
 \begin{align}\label{ge_mom}
   \bar{h}_{MN}(q) = \int \e^{iqx}\bar h_{MN}(x)
\,d^Dx = \frac{2\pi \vak m \, \delta(qu)}{q^2+i \varepsilon q^0}
\left(u_M u_N-\fr{1}{D-2}\,\eta_{MN}\right).
  \end{align}

The metric deviations due to the wall grow linearly with the
distance
 \begin{align} \label{brgr}
h_{MN}=\frac{\vak \mu}2\left(\Xi_{MN}
-\frac{D-1}{D-2}\,\eta_{MN}\right)|z|= \frac{2 k \hsp |z|}{\vak}
\:{\rm diag}\hsp(-1,1,...,1,{D-1})\,,
 \end{align}
where
 \begin{align} \label{k_def}
\Xi_{MN}=\Sigma_M^\mu \Sigma_N^\nu \eta_{\mu\nu}\,,\qquad k \equiv
\frac{\vak^2 \mu }{4(D-2)}\,,
 \end{align}
and the corresponding Fourier transform reads
 \begin{align} \label{brgrf}
h_{MN}(q)= \frac{(2\pi)^{D-1}{\vak \mu}}{q^2} \left(\Xi_{MN}
-\frac{D-1}{D-2}\,\eta_{MN}\right)\,.
 \end{align}

The first order correction to the particle motion $\delta z^M$ in
the field of the wall (\ref{brgr}) depends on the choice of the
parameter on the world line. Specifying it so that the deviation of
the einbein is zero,
\begin{align}
\label{e1eq} \delta e=-\frac{ m}{2} \left( \vak
h_{MN}u^Mu^N+2\,\eta_{MN}u^M \delta
 \dot{z}^N \vp \right)=0\;,
\end{align}
we obtain from the geodesic equation (\ref{eomp})
 \begin{align} \label{acce1}
 \delta\ddot{z}^{\,0} =2kv\,
\gamma^2 \;{\rm sgn }(\tau)\,,\qquad  \delta\ddot z \equiv
\ddot{z}^{D-1}=k\, (D\gamma^2 v^2+1)\;{\rm sgn } (\tau) \,,
 \end{align}
observing that the force is repulsive as expected. Integrating
(\ref{acce1}) twice with initial conditions $ \delta z^M(0)=0,
\;\delta\zt^M(0)=0$, one has
 \begin{align} \label{acce2}
 \delta z^0   =k v \tau^2\, \gamma^2
\;{\rm sgn }(\tau)\,,\qquad { \delta z }=\frac{1}{2}\, k\tau^2
\left(D\gamma^2 v^2+1\right)\,{\rm sgn }( \tau)\,.
 \end{align}
Substituting  (\ref{acce2}) back into (\ref{e1eq}) one can check
that the gauge condition $ \delta e=0$  holds indeed.

In order to find perturbations of the domain wall embedding
functions $\delta X^M$ due to gravitational interaction with the
particle one has to derive the linearized perturbation of the
Nambu-Goto equation specifying the world volume metric as an induced
metric
 \begin{align}
  \delta \gamma_{\mu\nu}=2\hsp \delta^M_{(\mu}\hsp \delta\nhsp \smhsp X^N_{\nu)}\eta_{MN}+
 \vak \bar h_{MN}\Sigma^{M}_{\mu}\Sigma^{N}_{\nu} \,,
 \end{align}
where brackets denote symmetrization over indices with the factor
$1/2$. Then linearizing the rest of  Eq. (\ref{em}), after some
rearrangements one obtains the following equation for deformation of
the wall:
 \begin{align}  \label{pisk}
 {\Pi}_{MN}\;\Box_{D-1}\;\delta\nhsp \smhsp
X^N= {\Pi}_{MN}\;J^N\,, \qquad  {\Pi}^{MN} \equiv \eta^{MN}-
\Sigma^{M}_{\mu}\Sigma^{N}_{\nu} \eta^{\mu\nu}\,,
 \end{align}
 where $\Box_{D-1} \equiv \partial_{\mu} \partial^{\mu}$ and $ {\Pi}^{MN} $ is the
projector onto the (one-dimensional) subspace orthogonal to
$\cV_{D-1}$. The source term   in  (\ref{pisk}) reads:
 \begin{align} \label{JN}
 J^N=   \vak \, \Sigma_P^\mu \,\Sigma_Q^\nu\,
\eta_{\mu\nu} \left(\frac{1}{2} \, \bar h^{PQ,N} -\bar
h^{NP,Q}\right)_{\!z=0}\! .
 \end{align}
Using the aligned coordinates on the brane
$\sigma^{\mu}=(t,\mathbf{r}), $ we will have
$\delta^{M}_{\mu}=\Sigma^{M}_{\mu}$, so the projector $ {\Pi}^{MN}$
reduces the system (\ref{pisk}) to a single equation for the $M=z$
component.   Generically, the transverse coordinates of the branes
can be viewed as Nambu-Goldstone bosons (branons) that appear as a
result of spontaneous breaking of the translational symmetry
\cite{KuYo}. These are coupled to gravity and matter on the brane in
the brane-world models  via the induced metric \cite{Bu}. In our
case of the codimension one, there is only one such branon. The
remaining components of the perturbation $\delta\nhsp \smhsp X^M$
can be removed by suitable transformation of the coordinates on the
world volume, so the equality $\delta\nhsp \smhsp X^\mu=0$ is
nothing but the choice of gauge. Note that in this gauge the
perturbation of the induced metric $\delta \gamma_{\mu\nu}$ does not
vanish, as it was for the perturbation of the particle einbein $e$.

 Denoting the physical component as $
\delta\nhsp \smhsp X^z\equiv\Phi(\sigma^\mu) $ we obtain the branon
$(D-1)$-dimensional wave equation:
 \begin{align} \label{NGEQ}
\Box_{D-1} \Phi(\si^{\mu})=J(\si^{\mu}),
  \end{align}
  with  the source term $J \equiv J^z$.
Substituting (\ref{hpart}) into the Eq.\,(\ref{JN}) we obtain
 \begin{align} \label{jxb}
 J( \sigma )=\vak \left(\frac{1}{2}\,
 \eta_{{\mu\nu}}\bar{h}^{\hsp \mu\nu,z}-\bar{h}^{\hsp  z\hsp  0,0}\right)_{\!z=0}
 =-  \fr{\la vt}{[\gamma^2 v^2
t^2+r^2]^\fr{D-1}{2}}\,,\end{align} with\begin{align} \la=\fr{\vak^2
m\gamma^2\Gamma\left(\fr{D-1}{2} \right)}{4\pi^{\fr{D-1}{2}}}\left(
\gamma^2v^2 +\fr{1}{D-2}\right) .
 \end{align}
The retarded solution to  Eq. (\ref{NGEQ}) consists of two parts
$\Phi= \Phi_{\ah}+\Phi_{\smhsp\bh}$, where the first is
antisymmetric in time and represents an eventual deformation of the
wall correlated with the particle motion.   The second part is the
spherical branon wave starting at the moment of perforation and
propagating to infinity with the velocity of light. This wave is not
the solution of the homogeneous branon equation, but is has a jump
at $t=0$ ensuring continuity of the full solutions. The explicit
expressions of  both parts were presented in \cite{GaMeS1}; they
depend on the dimension  of the space-time. Here we will not need
their explicit form, so we give only their integral representations
suitable for later use:
 \begin{align}\label{Phi11}
 \Phi_{\ah}
 &\equiv - \Lambda \,\sgn\nhsp(t)\,I_{\ah}\,,\qquad \Phi_{\smhsp\bh}
\equiv 2\, \Lambda  \, \theta(t)\,I_{\bh}\,,\qquad \Lambda \equiv
\frac{ \sqrt{\pi}
 \,\la}{2^{ \frac{D-2}{2}} \gamma^3\Gamma\left(\fr{D-1}{2}
 \right)}\,,\\
 \label{Phi13a}
  & I_{\ah}(t,r) = \frac{1}{r^{\fr{D-4}{2}}} \int\limits_{0}^{\infty} \! dk
\,J_{\frac{D-4}{2}}(k r)\,{k}^{\frac{D-6}{2}} \, \e^{- k\gamma
v|t|}\,, \\
 &I_{\bh}(t,r)= \frac{1}{r^{\fr{D-4}{2}}} \int\limits_{0}^{\infty} \!
dk\,J_{\frac{D-4}{2}}(k r)\,{k}^{\frac{D-6}{2}} \,\cos \smhsp kt\,
, \label{Phi13b}
 \end{align}
where $J_{\nu}{(z)}$ is a Bessel function of the first kind.

\section{Conservation of the energy-momentum}
In the first order in $\vak$ the total energy-momentum tensor
consists of three contributions (\ref{psi2eq}) and satisfies the
conservation equation (\ref{source22}). To convert the latter into
the the energy-momentum balance equation one has to integrate over
the world tube $\Omega$:
 \begin{align}
\label{conlaw} 0=\int_\Omega\partial_N\tau^{MN}=\int_{\partial
\Omega}\tau^{MN}d \Sigma_N\,,
 \end{align} bounded by the closed hypersurface
\begin{align}
\partial \Omega= \Sigma_{t_0}\cup
 \Sigma_{t_f} \cup \Sigma_{\infty}\,,
 \end{align}
consisting of two spacelike hypersurfaces associated with the
moments of time $t_0,\, t_f$ (usually chosen orthogonal to the time
axis), and the closing lateral hypersurface $\Sigma_{\infty}$ at
spatial infinity. To get the usual energy-momentum conservation
equation two conditions should hold: i) finiteness of the  the
integral of $\tau^{MN}$ over $\Sigma_{t}$ which is interpreted as
the $D$ -momentum vector,
\begin{align}\label{momendef}
 P_{\rm tot}^M(t)=\int_{\Sigma_{t}} \tau^{MN}{\kern 1pt}d \Sigma_N=  \int  \tau^{M0}     dz\, d^{D-2}
\mathbf{r}\,,
 \end{align}
and ii) vanishing of the lateral flux $\tau^{MN}$ through the
timelike hypersurface $\Sigma_{\infty}$. This is usually guaranteed
by the sufficient  falloff  of the integrand at infinity. In the
case of the domain wall both conditions are not satisfied. First,
the wall is considered an infinite and having finite mass density,
so the total energy in the zero order in $\vak$ diverges. We will
see shortly that the corresponding contribution diverges also in the
linear in $\vak$ order. Secondly, the lateral flux for the wall is
nonzero since the integrand does not fall fast enough at spatial
infinity. So in our case the momentum equation reads
\begin{align}\label{pof}
P_{\rm tot}^M(t_f)-P_{\rm tot}^M(t_0)=  -\int_{\Sigma_{\infty}}
\tau^{MN}d \Sigma_N\,.
 \end{align}
According to the split of the total energy-momentum tensor
(\ref{source22}) we can write
\begin{align}\label{momensplit}
 &P_{\rm tot}^M(t)=\delta\bar{P}^M(t)+\delta P^M(t)+S^M(t)
 \,, \end{align} where
 \begin{align}\label{momenp}
&\delta\bar{P}^M(t)=
 \int \delta\bar{T}^{M0}\,  dz\, d^{D-2}\mathbf{r}\,,\\\label{momenw}
& \delta P^M(t)=  \int  \delta T^{M0}  \,  dz\,
 d^{D-2}\mathbf{r} \end{align}
are the first-order kinetic  momenta carried by the particle
\footnote{As we have already noted, the particle kinetic momenta
defined as the integral (\ref{momenp}) does not coincide with the
generalized Hamiltonian momentum $m {\dot z}^Ng_{MN}$ once gravity
is taken into account. Our present definition, however, is more
convenient for the further analysis.}
 and the wall, while
 \begin{align}\label{momens}
  S^M(t)=  \int  \delta S^{M0}  \,  dz\, d^{D-2} \mathbf{r}\, \end{align}
 is the momentum carried by their
gravitational field. The lateral flux at the right-hand side of Eq.
(\ref{pof}) can also be split into three similar  contributions. The
boundary hypersurface $\Sigma_{\infty}$ in the $(D-1)$-dimensional
space consists  of three components
\begin{align}
\Sigma_{\infty}=  T\times\left(B_-\cup B_+  \cup D_R\right)\,,
 \end{align}
where $T$ is the time real axis, and  $B_\pm\,,D_R$ are  the
$(D-2)$-dimensional surfaces: $B_\pm=\{{\rm all}\; {\bf{r}}\,,
z\to\pm\infty\}$ (with an
 associated measure $d^{D-2}{\bf{r}}$) and $D_R=\{ {\rm all}\; z\,,
 R=|{\bf{r}}|\to \infty\}$ (with the measure
 $R^{D-3}d^{D-3}\Omega$). Actually, all the fluxes through $B_{\pm}$ vanish,
as well as the fluxes of $\delta\bar{T}^{Mr}\,,\delta S^{Mr}$
through $D_R$, but not the flux $\delta T^{Mr}$ representing the
contribution of the wall. We are therefore
  left with
 \begin{align}\label{momf}
P_{\rm tot}^M(t_f)-P_{\rm tot}^M(t_0) =-\lim_{R\to
\infty}\left(\int_{t_0}^{t_f}dt\int_{-\infty}^{\infty}\,dz\int_{S^{D-3}_{R}}
\delta T^{Mr}R^{D-3}d^{D-3}\Omega\right)\,.
 \end{align}
 Thus the difference between the momenta defined in a standard way as
the integrals over spacelike hypersurfaces  is related to some
integral over the corresponding time interval. Another unpleasant
feature is that the integrals (\ref{momenw}) are divergent for an
infinite wall. To cure both of these drawbacks one could introduce
the cutoff volume for the wall, but this make the analysis more
complex. Instead we pass to time derivatives of the momenta which
are all finite. In other words we check the momentum conservation
between the infinitely close moments of time. Then the Eq.
(\ref{momf}) gives
\begin{align}\label{difcons}
 \frac{d}{dt}\left(\delta\bar{P}^M(t)+\delta P^M(t)+S^M(t)\right)= - \lim_{R\to
\infty}\biggl(\int_{-\infty}^{\infty}\, dz\int_{S^{D-3}_{R}}\delta
T^{Mr}R^{D-3}d^{D-3}\Omega\biggr)\equiv f^M\,,
\end{align}
where the integral at the right-hand side will be called the lateral
momentum flux. This term looks like an external force acting upon
the system, but in fact it is due to an additional loss of the wall
momentum. In principle it could be absorbed by the redefinition of
the wall  momentum at the left-hand side, but we prefer to keep the
usual definition (\ref{momenw}).

\section{Computation of the momenta}
We proceed in analyzing various contributions to the differential
conservation equation (\ref{difcons}). Note that in the {\em zero}
order in $\vak$ the particle and the wall kinetic momenta   are
simply $$ \bar{P}^M =m u^M\,,\qquad {P}^{M}= \mu
V_{\br}\delta^M_0\,,$$ where $V_{\br}$ is the world volume
introduced for normalization. These quantities are constant which
can be omitted from  further analysis.
\subsection{Kinetic momenta}
The first-order   particle stress tensor is obtained  expanding the
general expression (\ref{EMTp}) in $\vak$:
\begin{align}
\label{T1MN_r}  \delta \bar {T}^{MN}(x)= \frac{m}{2} \int \left[\hsp
4 \hsp  \delta\dot z^{(M} u^{N)} - \, u^M u^N \left( \vak {h}+ 2
\hsp \delta{z}^P\pa_P \vp \right) \right] \delta^D\!\left(x- u
\tau\right)\, d\tau\,,
\end{align}
where $h $ is the trace of the first-order metric deviation due to
the wall (\ref{brgr}); the symmetrization over the indices $(MN)$ as
well as the antisymmetrization $[MN]$  below is defined  with 1/2.
The delta function indicates the localization of the integrand at
the nonperturbed particle world line. (Note that our integral
definition of the  kinetic momentum coincides with the Hamiltonian
definition of the covariant generalized momentum
$P_{M}^{(h)}=\partial L/\partial \dot z^m$ only in the zero order in
the gravitational constant.)

The first-order stress tensor of the wall is obtained  substituting
the first-order metric deviation (\ref{hpart}) due to the particle
and the first-order perturbations of the wall world volume into Eq.
(\ref{EMTw}):
\begin{align} \label{taumn}
   \delta T^{ {M} {N}}  (x) =
\fr{\mu}{2} \int & \left[ 4\, \delta_{\mu}^{(M}\delta{\nhsp\smhsp
X}_{\vm{\mu} \nu}^{N)} \eta^{\mu\nu}   -2\, \delta^M_\mu
\delta^N_{\nu\vm{\mu}} \left( \vak \bar{h}^{ \mu\nu}  + 2\,
\eta^{LR} \delta_R^{( \mu } \delta{\nhsp\smhsp X}_L^{\nu)} \right) +
\right. \nn\\ & \left. \;\;+
 \delta^M_{\mu}\delta^{N}_{\nu\vm{\mu}} \eta^{ \mu\nu} \left(\vak \bar{h}^{\lambda}_{\lambda}-
\vak \bar{h}+2\hsp \delta{\nhsp\smhsp
X}^{L}_{\lambda}\delta^{\lambda}_L -2\hsp \delta{\nhsp\smhsp X}^{L}
\partial_L \vp\right) \right]\delta^{D-1}\!\left(x -  \sigma \right)\,\delta(z)\: d^{D-1}\si
\,.
 \end{align}
Again, the delta functions in the integrand indicate   its
localization on the unperturbed wall  world volume.

 Due to the
kinematics of the collision, the {\em first-order} kinetic momenta
also have nonzero only the $0$ and $z$ components. The particle
momentum is calculated substituting the wall  metric deviation
$h_{MN}(\tau)$ given by (\ref{brgr}) and the particle world line
deviation $ \delta z^M(\tau)$ given by (\ref{acce2}) into
(\ref{T1MN_r}) and integrating with the help of the delta function:
\begin{align} \label{P_par}
   \delta \bar{P}^{z}  = m k
 \left[(3D-2)\ga v^2+\ga^{-1}\vp \right] |t| \,, \qquad   \delta\bar{P}^{0}  =2D m k  \ga v |t| \,.
 \end{align}

Now calculate the time component of the wall momentum. Substituting
the deviation $\delta X^M=\delta^M_z \Phi$ with $\Phi$ given by
(\ref{Phi11}) into the integrand of (\ref{momenw}) we get
\begin{align} \label{taumn3r}
    \delta  T^{{0}{0}}  =\fr{\mu}{2}  \left[\left( -2\hsp \vak \bar{h}_{00}   +
\vak  \bar{h}_{zz} \vp\right) \delta(z) -2\,\Phi\,
 \delta'(z)\right]  \,  .
 \end{align}
Since $\Phi$ is the function of the world-volume coordinates $(t,r)$
only, the term $\Phi \,
 \delta'(z)$ vanishes upon integration over $z$, so $\delta P^0$ does not depend on $\Phi$.
Substituting into the second quantity the particle metric deviation
(\ref{hpart}) one gets
\begin{align} & \delta T^{00}=  \frac{\Gamma\left(\fr{D-3}{2}\right)}{2\pi^{\fr{D-1}{2}}}
\left(\vp(D-2)\gamma^2 v^2 +(2D-7) \right)  m k \chi\, \delta(z) \,
, \\&\label{chi} \chi \equiv \frac{1}{\left[\gamma^2(z-v
t)^2+r^2\right]^{\fr{D-3}{2}}}\,,
\end{align}
so the first-order zero component will read
\begin{align}  \delta P^0=  \frac{\Gamma\left(\fr{D-3}{2}\right)}{\sqrt{\pi}\Gamma\left(\fr{D-2}{2}\right)}
\left(\vp(D-2)\gamma^2 v^2 +(2D-7) \right)  m k Q\,,
\end{align}
where the  integral of $\chi$ over $r$ including the volume factor
\begin{align} \label{Q} Q(a) = \int_0^{\infty} \frac{r^{D-3}\; dr}{ \left(a^2+r^2\right)^{ \frac{D-3}{2}}
}\,,
 \end{align}
with $a^2=\gamma^2(z-v t)^2$, linearly diverges at the upper limit.
This is not surprising taking into account an infinite extension of
the wall. To avoid a cumbersome normalization procedure, we pass
from the momentum to its time derivative. This will be sufficient to
define gravitational dressing of the kinetic momenta which is our
main goal here. The derivative $\dot Q$ is finite and the
corresponding integral is easily evaluated by the substitution $
1+(r/a)^2=1/y$ leading to Euler's beta function:
\begin{align}\label{cucumber0}   \int \frac{r^{D-3}\; dr}
{ \left(a^2+r^2\right)^{ \frac{D-1}{2}} }=
\frac1{|a|}\frac{\sqrt{\pi}\hsp
 \Gamma\left(\fr{D-2}{2}\right)}{ 2\hsp \Gamma\left(\fr{D-1}{2}\right)}
 \,.
 \end{align}
Since the unperturbed momentum of the wall is constant (also
infinite), we can interpret the resulting quantity as describing the
derivative of the full momentum up to the first order simply by
omitting  $\delta$:
 \begin{align} \label{muchomort3}
  \dot{P}^0=- \gamma v
\left(\vp(D-2)\gamma^2 v^2 + 2D-7  \right) m k
 \,\sgn(t)
    \,.
 \end{align}

The computation of the spatial component of the wall momentum is
more involved. The flux $T^{z0}$  can be simplified as follows:
 \begin{align} \label{taumn3z}
     \delta T^{0z}  =
 {\mu}  \hsp  \Phi_{,\smhsp 0}(t, {r})\,  \delta(z)\,
,
 \end{align}  where the wall perturbation
 is the sum of two terms $\Phi=\Phi_{\ah}+\Phi_{\bh}$, the first describing
the regular deformation induced by the particle gravitational field,
and the second corresponding to the shock branon wave emerging at
the moment of piercing and then freely propagates outwards along the
wall. Substituting (\ref{Phi11}) into (\ref{taumn3z}) and taking
into account that $I_{\ah}|_{t=0}=I_{\bh}|_{t=0}$ \footnote{This
follows from (\ref{Phi13a},\,\ref{Phi13b}) and is explained in
detail in Eq. (5.39) of \cite{GaMeS1}.}, one can verify the absence
of terms proportional to $\delta(t)$ in the time derivative of the
total perturbation $\Phi_{,0}$. Thus one can write $\delta P^z(t) =
\delta P^z_{\ah}(t)+ \delta P^z_{\bh}(t)$ with
 \begin{align} \label{taumn3r_1}
  \delta P^z_{\ah}=-\Lambda \hsp{\mu}\, \sgn (t)
  \int  I_{\ah,\smhsp 0}(t, {r})\: d^{D-2} \mathbf{r}\,,
 \qquad  \delta P^z_{\bh}=
 2 \,\Lambda \hsp {\mu}\,
\theta(t)\, \int I_{\bh,\smhsp 0}(t, {r})\: d^{D-2} \mathbf{r}\,.
 \end{align}
Let us start with  the "regular" part $ P^z_{\ah}$. Substituting
$I_{\ah}$  (\ref{Phi13a})  and  performing an integration over the
sphere we obtain:
 \begin{align} \label{taumn3r_2}
  \delta P^z_{\ah}=
 \frac{2\hsp \pi^{\frac{D-2}{2} } \mu   \Lambda \gamma v}{ \Gamma\left(\frac{D-2}{2} \right) } \int
\,J_{\frac{D-4}{2}}(k r)\,{k}^{\frac{D-4}{2}} \, \e^{- k\gamma v|t|}
r^{\frac{D-2}{2}}\:   dk\,dr\,.
 \end{align}
Using the integral
 \begin{align} \label{taumn3r_3}
  \int\limits_0^{\infty}
 J_{m}(k r)\,{k}^{m} \, \e^{-ak}  \,   dk=\frac{(2r)^m \Gamma\!\left( m+ \frac{1}{2}\right)}
 {\sqrt{\pi} \left(a^2+r^2\right)^{m+1/2}}\,,
 \end{align}
 one obtains again the  divergent quantity
 \begin{align} \label{taumn3r_4}
 \delta P^z_{\ah}= \frac{mkv }{\sqrt{\pi}}
  \frac{ \Gamma\left(\frac{D-3}{2} \right) }{
\Gamma\left(\frac{D-2}{2} \right) }\left(\vp (D-2)\gamma^2 v^2 +1
\right) Q(a)\, ,
 \end{align}
 now with $a^2=\gamma^2 v^2 t^2$. Passing  to the time
derivative we use the fact that $\delta(t) \hsp\dot{I}_{\ah}=0 $ in
the distributional sense. Taking into account that there is no
zero-order contribution to $P^z$, we can write
 \begin{align} \label{muchomort2}
 \dot{P}^z_{\ah}=-  mk\left(\vp (D-2)\gamma^2 v^2 +1 \right)\gamma
v^2 \,\sgn(t)
    \,.
 \end{align}

Now we present the corresponding quantities for the particle,
differentiating the sum of the zero and the first-order
(\ref{P_par}) momenta:
\begin{align} \label{f_par}
 \bar{F}^z \equiv \dot{\bar{P}}^{z}  = m k
 \left[(3D-2)\ga v^2+\ga^{-1}\vp \right] \sgn(t) \,, \qquad
  \bar{F}^0 \equiv\dot{\bar{P}}^{0}  =2D m k  \ga v \,\sgn(t) \,.
 \end{align}
  All the momenta
derivatives (\ref{muchomort3}), (\ref{muchomort2}), and
(\ref{f_par}) are constant before and after the  moment of piercing
$t=0$ when they change the sign.  The sum $\dot P^M+\dot{\bar{P}}^M$
does not vanish for both values of $M$. This is not surprising since
we still need to add contribution of the gravitational stresses.

\subsection{Branon contribution}\label{branon1}
One can check that the shock wave (branon)  part of the wall's
perturbation $\Phi_{\bh}$ does not give contribution to the zero
component of the momentum (the energy). However there is still the
branon contribution to  $P^z$  arising  after the perforation.
Substituting the integral representation (\ref{Phi13b}) for
$I_{\bh}$ into the Eq. (\ref{taumn3r_1}), one obtains
 \begin{align} \label{yhn1}
\delta P^z_{\bh}= -
 2 \,\Lambda \hsp {\mu} \Omega_{D-3}\,
\theta(t)\, \int   k^{\frac{D-4}{2}} J_{\frac{D-4}{2}}(kr)\,\sin kt
\,r^{\frac{D-2}{2}} \,  dk dr\,.
 \end{align}
Integration over $k$ is performed using the integral \cite{GR}
$$ \int\limits_0^{\infty}   k^{\nu} J_{\nu}(kr)\,\sin
kt \,  dk=  \frac{\sqrt{\pi}\hsp
(2r)^{\nu}(t^2-r^2)^{-(\nu+1/2)}}{\Gamma\left(1/2-\nu
\right)}\,\theta(|t|-r)\,,$$ to yield
 \begin{align} \label{yhn_2a}
\delta P^z_{\bh}= -
  \frac{  2
^{\frac{D }{2}} \pi^{\frac{D-1}{2}}}{ \Gamma\!
\left(\frac{D-2}{2}\right) \Gamma\left(- \frac{D-5}{2} \right)}\,
\Lambda \hsp {\mu}  \, \theta(t)\, \int (t^2-r^2)^{-(D-3)/2}
\,\theta(|t|-r)  \,r^{D-3} \, dr\,.
 \end{align}
The latter expression contains $\Gamma\left(- \frac{D-5}{2} \right)$
which has a simple pole at odd $D \geqslant 5$. Thus (\ref{yhn_2a})
drastically depends upon the parity of $D$.

First let us consider odd $D\geqslant 5$. Applying the
distributional limit [Sec. 3.5, Eq.\,(1) of \cite{Gelf}].
 \begin{align} \label{yhn_2b}
 \lim_{\lambda \to -n} \frac{ [x \, \theta(x)]^{\lambda-1} }{\Gamma (\lambda )} = \frac{d^n}{dx^n}\,\delta (x)\equiv \delta^{(n)} (x)\,,
  \end{align}
one gets
 \begin{align}\label{yhn_2c}
\delta P^z_{\bh} & =
 -\frac{2^{{D}/{2}}  \pi^{\frac{D-1}{2}}}{ \Gamma\! \left(\frac{D-2}{2}\right) }\,\Lambda \hsp {\mu}\,
\theta(t)\, \int  \delta^{\left(\frac{D-5}{2}\right)}(t^2-r^2)
\,r^{D-3} \, dr\nn
\\ & =-
 \frac{2^{\frac{D-2}{2}}  \pi^{\frac{D-1}{2}}}{ \Gamma\! \left(\frac{D-2}{2}\right) }\,\Lambda \hsp {\mu}\,
\theta(t)\,  \left(\frac{\partial}{\partial t^2}
\right)^{\!\frac{D-5}{2}}\!\!  \int \delta(t-r) \,r^{D-4} \, dr\,,
\end{align}
where the order of  derivative is integer and we pass to the
differentiation over $t^2$.
 Integrating trivially over $r$ and next differentiating with respect to $|t|^2$ according to
 \begin{align}\label{yhn_2d}
 \frac{d^{\lambda}}{dx^{\lambda}}\: x^{\rho}= \frac{\Gamma(\rho+1)}{\Gamma(\rho-\lambda+1)}\: x^{\rho-\lambda}\,,
\end{align}
easily verified for integer $\lambda$, one obtains finally:
 \begin{align} \label{yhn3}
  \delta P^z_{\bh}   =-
  \fr{\vak^2 \hsp {\mu} m  }{2 \hsp \gamma }\left( \gamma^2v^2
+\fr{1}{D-2}\right) t\,\theta(t)\,.
 \end{align}
The corresponding force $F^z_{\bh}=\delta \dot{P}^z_{\bh}$  reads
 \begin{align} \label{yhn4}
    F^z_{\bh} = -
  \fr{2 \hsp k m  }{\gamma }\left( (D-2)\,\gamma^2v^2+
1\vp \right)  \theta(t) \,.
 \end{align}

Now consider even $D\geqslant 4$: the gamma function is regular now
and we represent
 \begin{align} \label{yhn4_a}
 \frac{(t^2-r^2)^{-(D-3)/2}}{\Gamma\left(- \frac{D-5}{2} \right)}=\frac{1}{\sqrt{\pi}}\left(\frac{\partial}{\partial t^2}
\right)^{\!\frac{D-4}{2}} \frac{1}{\sqrt{t^2-r^2}}\,.
 \end{align}
Substituting (\ref{yhn4_a}) into (\ref{yhn_2a}) one obtains
 \begin{align} \label{yhn4_b}
\delta P^z_{\bh}= -
  \frac{  2
^{\frac{D }{2}} \pi^{\frac{D-2}{2}}}{ \Gamma\!
\left(\frac{D-2}{2}\right)  }\, \Lambda \hsp {\mu}  \,
\theta(t)\,\left(\frac{\partial}{\partial t^2}
\right)^{\!\frac{D-4}{2}} \int_0^t  \frac{ r^{D-3}}{\sqrt{t^2-r^2}}
\, dr\,.
 \end{align}
The variable change $y=r^2/t^2$ leads us again to the beta function
$\mathrm{B}\left(\frac{D-2}{2},\frac{1}{2}  \right)$ so
 \begin{align} \label{yhn4_c}
\delta P^z_{\bh}= -
  \frac{  2
^{\frac{D-2 }{2}} \pi^{\frac{D-1}{2}}}{ \Gamma\!
\left(\frac{D-1}{2}\right)  }\, \Lambda \hsp {\mu}  \,
\theta(t)\,\left(\frac{\partial}{\partial t^2}
\right)^{\!\frac{D-4}{2}}t^{D-3}\,.
 \end{align}
Applying (\ref{yhn_2d}) for integer $\frac{D-4}{2}$, the momentum
carrying by branon reads
 \begin{align} \label{yhn4_d}
\delta P^z_{\bh}= -
     2\hsp (2\pi)^{\frac{D-2}{2}} \, \Lambda \hsp {\mu}  \,
\theta(t)\,t\,.
 \end{align}
Substituting $\Lambda$ (\ref{Phi11}), one arrives at the same
expression (\ref{yhn3}), as for odd space-time
dimensionality\hsp\footnote{Not surprisingly,  we have the same
formula. In fact, the convolution of the test function $\varphi \in
C^{\infty}$ with $\delta^{(n)} (x)$, $n\in \mathbb{N}$ returns its
$n$th derivative; hence according to (\ref{yhn_2b}), the convolution
with the analytic functional $[x \, \theta(x)]^{-(\lambda+1)}
/\Gamma (-\lambda )$ can be regarded as defining the fractional
derivative of order $\lambda$. With this definition the differential
property (\ref{yhn_2d}) becomes well defined and valid for
\emph{any} $\lambda \in \mathbb{R}$. The same concerns semi-integer
derivatives of $\delta(t^2-r^2)$ in (\ref{yhn_2c}). Thus in the
sense of fractional derivatives of distributions, these two ways to
derive Eq. (\ref{yhn3}) are equivalent.}. Notice that the final
result keeps this form also for $D=2,3$.

This gives rise to another problem: while all  other contributions
to momenta transfer are proportional to sign functions of time,
(\ref{yhn4}) is proportional to the Heaviside function. Since
$\sgn(t)$ and $ \theta(t)$ are linearly independent, this extra
contribution cannot fix the above nonconservation problem. We will
see shortly, that this time-asymmetric part is related to the
nonzero lateral flux of the momentum.

\subsection{Gravitational stresses}
We start by analyzing the component $S^{z0}(h,\bar h)$ obtained by
substituting the metric deviations $h_{MN}$ (\ref{brgr}) and
$\bar{h}_{MN}$ (\ref{hpart}) into  (\ref{natag_0}). After
rearrangements one obtains nonzero contributions of two types:
  \begin{itemize}
    \item the first derivatives of both $h_{MN}$ and $\bar{h}_{MN}$.
    Using
     the fact that $\bar{h}_{MN,0}=-v \bar{h}_{MN,z}$ and $\bar{h}_{MN,00}=v^2
    \bar{h}_{MN,zz}$,
this contribution reduces to
$$  \fr{
 mkv\,\Gamma\left(\fr{D-3}{2}\right)}{4\pi^{\fr{D-1}{2}}} \left(\vp
(D-2)\gamma^2 v^2 +3 \right)\chi_{,z}\; \sgn (z)
$$ Integrating over $z$ by parts, one obtains $\delta(z)$ showing that it is localized on the
wall.
    \item Terms containing the box operator acting on $\bar{h}_{MN}$, namely, $$\left(-2 h^{00}+h^{zz}\right)\Box
    \bar{h}^{0z}\,.$$ Using the first-order Einstein equation for
    $\bar{h}^{MN}$  $$  \Box \bar{h}^{MN}=-\vak\! \left(
\bar{T}^{MN}-\frac{1}{D-2}\, \bar{T}\,\eta^{MN} \right) \,, \nn
 $$ one can see that these are localized on the particle's
 world line.
\end{itemize}
So apparently nonlocal stresses localize on the wall and the
particle world volumes. From the above calculations it is clear that
it happens because one deals with the products of two Coulomb-like
fields which are tight to the sources without the retardation. Thus
we present the integral of  $S^{z0}$ as the sum  $\bar S^{z }$ and $
S^{z }$ according to their localization:
 \begin{align} \label{Ssplit}
 \int S^{z0}  dz d^{D-2} \mathbf{r}=S^z+ \bar S^z\,,
\end{align}
where explicitly
 \begin{align} \label{Ssplit1}
&  \bar S^z= \vak\mu\gamma v\int\left( h^{zz}-2
h^{00}\right)\delta(z-vt)\; dz \,d^{D-2} \mathbf{r}\,,\\& S^z=\fr{
 mkv\,\Gamma\left(\fr{D-3}{2}\right)}{4\pi^{\fr{D-1}{2}}} \left(\vp
(D-2)\gamma^2 v^2 +3 \right)\int \chi_{,z}\; \sgn (z) \;  dz\,
d^{D-2} \mathbf{r}\,,
\end{align}
with $\chi$ defined by (\ref{chi}):

$$\bar{S}^{z0} =\left(-2
h^{00}+h^{zz}\right)\Box
    \bar{h}^{z0} \,, \qquad   {S}^{z0} =  \fr{
 mkv\,\Gamma\left(\fr{D-3}{2}\right)}{4\pi^{\fr{D-1}{2}}} \left(\vp
(D-2)\gamma^2 v^2 +3 \right)\chi_{,z}\; \sgn (z)\,. $$ Integrating
$\bar{S}^{0z}$ over $z$ and $r$, we obtain the finite contribution
to the particle momentum,
 \begin{align} \label{muchomort3B}
 {\bar{S}}^z=-2 \left(D+1\right)m k \gamma v^2 \,|t|\,,
 \end{align}
 and the corresponding time derivative is
 \begin{align} \label{muchomort3A}
 \dot{\bar S}^z = -2 \left(D+1\right)m k \gamma v^2 \,\sgn
(t)\,.
 \end{align}
The second integral is evaluated by integration by parts over $z$
and then using the arising delta function. The radial integral
diverges as before:
 \begin{align} \label{taumn3r_5}
S^{z}& =- \fr{
 mkv\,\Gamma\left(\fr{D-3}{2}\right)}{ \sqrt{\pi}\hsp \Gamma\left(\fr{D-2}{2}\right)} \left(\vp
(D-2)\gamma^2 v^2 +3 \right)\int r^{D-3}\left.\chi
\vphantom{d_d^d}\vp\right|_{z=0} dr=\\&=- \fr{
 mkv\,\Gamma\left(\fr{D-3}{2}\right)}{ \sqrt{\pi} \hsp\Gamma\left(\fr{D-2}{2}\right)} \left(\vp
(D-2)\gamma^2 v^2 +3 \right) Q\,.
 \end{align}
The corresponding derivative is finite:
 \begin{align} \label{muchomort4}
 {f}^z \equiv \delta \dot{P}^z_{S} = \ga v^2
 \left[(D-2)\ga^2v^2+3\vp \right] m k \,\sgn(t)
    \,.
 \end{align}

Now consider the $S^{00}$ component. The following contributions are
nonzero:
\begin{itemize}
    \item Terms with first derivatives of both $h_{MN}$ and $\bar{h}_{MN}$:
$$ \frac{\Gamma\left(\fr{D-3}{2}\right)}{4\pi^{\fr{D-1}{2}}}
\left[\ds\vp (D-2)\gamma^2 v^2 +5 \right]\chi_{,z}\, mk \, \sgn
(z)\, . $$ These are localized to the wall integrating by parts over
$z$.
\item Terms with the second $z$ derivatives of $\bar{h}_{MN}$:
$$
\frac{\Gamma\left(\fr{D-1}{2}\right)}{\pi^{\fr{D-1}{2}}\gamma^2}\,m
k  |z| \chi_{,zz} \,.$$ These are localized to the wall integrating
over $z$ by parts twice .
\item Second derivatives of $ {h}_{MN}$  are  directly localized on the wall:
$$-\frac{\Gamma\left(\fr{D-3}{2}\right)}{ \pi^{\fr{D-1}{2}}}\,
(D-5)\,m k \,\delta(z) \chi \,.$$
    \item Boxes of $\bar{h}_{MN}$: $$\bar{S}^{00} \equiv -3 \hsp h_{00} \,\Box \bar{h}_{00}+\frac{1}{2}\, h_{00}\,
     \Box \bar{h}+ h_{PQ} \,\Box \bar{h}^{PQ}\,.  $$ These are
     localized on the particle world line after application of
     linearized Einstein equations.
\end{itemize}
Thus the last contribution gives the finite energy
 \begin{align} \label{muchomort51}
  {\bar{S}}^0 =-2 \left(\vp  \left(D+1\right) v^2 +\fr{4}{\ga^2}\right)
    m k\ga v |t|\,,
 \end{align}
 the corresponding derivative being
\begin{align} \label{muchomort5}
  \dot{\bar{S}}^0= -2 \left(\vp  \left(D+1\right) v^2 +\fr{4}{\ga^2}\right)
    m k\ga v \,\sgn (t)
    \,.
 \end{align}
The first three contributions attributed to the wall are integrated
exactly as before leading to divergent total energy, but finite time
derivative
 \begin{align} \label{muchomort6}
 \dot{ {S}}^0=m k\ga v \left[(D-2)\ga^2v^2+2D-5- \frac{2(D-3)}{\ga^{2}}\right] \,\sgn (t)
    \,.
 \end{align}
\section{Gravitational dressing}
Let us briefly summarize basic features of the particle-wall
piercing collision in the perturbative approach. In zeroth order in
gravitational coupling $\vak$ the wall is plane, unexcited, and
extending to spatial infinity. Its total momentum is constant and
infinite. The particle is moving with the constant velocity
orthogonally to the wall, its momentum is constant and finite.
Gravitational interaction between them is repulsive and causes
deceleration of the particle before the moment of perforation at
$t=0$ and acceleration after the perforation. The perturbation of
the particle world line is strictly time antisymmetric. The action
of the particle gravity upon the wall is more complicated:  the
wall's deformation consists of the time antisymmetric component due
to continuously varying gravitational force and a shock-wave
component that arises after the perforation.

In the first order in $\vak$ the total conserved (in Minkowskian
sense) energy-momentum tensor consists of three parts: two kinetic
terms and the stress tensor of the gravitational field. Since the
latter is constructed from the metric deviations generated by
time-independent sources, the presumably nonlocal gravitational
stresses in fact localize at the unperturbed particle's world line
and the wall's word volume. Therefore we can associate the
corresponding gravity contributions with kinetic terms obtaining
``dressed'' momenta of the particle and the wall. The associated
integrated total momenta are infinite due to slow falloff of
deformations at spatial infinity. To get rid of infinities we passed
to time derivatives of momenta, which actually represent the total
forces acting on the particle and the wall. These latter are finite
and we can explore the energy-momentum balance in the form of the
third Newton's law. Note that the contribution of the shock wave
makes the balance nonsymmetric in time. This contribution, however,
applies only to to spatial component of momentum, and does not
influence the energy balance.

Now we show that one can  construct the gravitationally dressed
momenta of the particle and the wall such that the total momentum
(\ref{momensplit}) satisfying the balance equation (\ref{difcons})
be the sum of two but not three quantities
 \begin{align} \label{dress}
 P^M_{\rm tot}=\bar{\cal{P}}^M+{\cal{P}}^M \,.
 \end{align}
For this it is enough to split the contribution of gravitational
stresses between the particle and the wall according to their
localization revealed in the previous section.
\subsection{Dressed particle momentum}
Both kinetic and gravitational contributions to the particle
momentum are finite, so we introduce the total dressed momentum as
the sum
 \begin{align} \label{muchomort6b}
 \bar{\cal{P}}^M= \delta \bar{P}^{M} + \bar{S}^{M}
    \,.
 \end{align}
 Substituting here (\ref{P_par}) and  (\ref{muchomort3B}) we obtain
the following nonzero components:
 \begin{align} \label{muchomort6c}
 &\bar{\cal{P}}^0=2\left[(D-3)(1-v^2)-1\right]kvm\gamma|t|\,,\\
  &\bar{\cal{P}}^z=2\left[(D-5){\kern 1pt}v^2+1\right]k m\gamma|t|\,.
 \end{align}
Note that the gravitational stresses contribution to the energy
(\ref{muchomort3B}) is negative, so the total first-order
 contribution to the  dressed energy may have negative sign depending
on the particle velocity.
\subsection{Dressed wall momentum}
 For the wall we write similarly
 \begin{align} \label{drew}
  {\cal{P}}^M=   \delta {P}^{M} + {S}^{M}
    \,,
 \end{align}
where the kinetic contribution  consists  of the sum of the regular
and the branon parts $\delta {P}^{M}=\delta {P}_{\ah}^{M}+\delta
{P}_{\bh}^{M}$. Actually the branon part $\delta {P}_{\bh}^{M}$ is
nonzero only for the spatial component $M=z$, while for the time
component we have
\begin{align} \label{drew1}
  {\cal{P}}^0=   \delta {P}_{\ah}^{{\kern 1pt}0} + {S}^{0}=\frac{\Gamma\left(\fr{D-3}{2}\right)}
  {\sqrt{\pi}{\kern 1pt}\Gamma\left(\fr{D-2}{2}\right)}\left[ (D-3)(1-v^2)-1
  \right]2{\kern 1pt}mk{\kern 1pt} Q(a)
    \,,
 \end{align}
with $Q(a)$ given by (\ref{Q}) with $a^2=\gamma^2 v^2 t^2$. This is
a divergent quantity, but its time derivative is finite. Using  Eq.
(\ref{cucumber0})  it  is easy to establish the identity
\begin{align} \label{drewss}
  \frac{d}{dt}{\cal{P}}^0=-\frac{d}{dt}{\bar{\cal{P}}}^0
    \,,
 \end{align}
showing that the change of the dressed wall's energy per unit time
is opposite to the change of the dressed particle's energy.

For the spatial component ${\cal{P}}^M$ we have two complications.
First, the shock wave contribution $\delta {P}_{\bh}^{M}$ is
nonzero. Second, for this component the lateral flux of momentum is
also nonzero. It turns out, that the regular and the branon parts as
the functions of time are linearly independent, so the balance
equation (\ref{difcons}) must hold for them separately. So consider
first the regular kinetic part $\delta {P}_{\ah}^{z}$. Summing up
the expressions (\ref{taumn3r_4}) and (\ref{taumn3r_5}) one gets
\begin{align} \label{drew2}
  {\cal{P}}^z=   \delta {P}_{\ah}^{z} + {S}^{z}=
  \frac{ \Gamma\left(\frac{D-3}{2} \right) }{\sqrt{\pi}
\Gamma\left(\frac{D-2}{2} \right) }\;2{\kern 1pt}mkv {\kern
1pt}Q(a)\,.
 \end{align}
Computing the time derivative of the difference between the wall and
the particle momenta we find
\begin{align} \label{drew3}
  \frac{d}{dt}{\cal{P}}^z=-\frac{d}{dt}{\bar{\cal{P}}}^z +k m\gamma\,\sgn(t)\left[(D-3) v^2+1\right]
    \,.
 \end{align}

\subsection{The lateral flux}
The origin of the extra force at the right-hand side of
(\ref{drew3}) lies in the nonzero flux of the $z$ component of the
brane kinetic momentum through the lateral boundary of the world
tube in accordance with (\ref{difcons}).
 After the routine consideration of all
components of the energy-momentum tensor, only one contribution of
the lateral flux survives, namely, the flux of the wall's $\delta
T^{zr}$ over $dS_{r}=r^{D-3}\Omega_{D-3} \,dr \hsp dt$. We obtain:
\begin{align}
{\sf f}^{z}\equiv \frac{d}{dt}\int { T}^{zr} dS_{r}=-
{\mu}\Omega_{D-3} \lim_{r \to \infty}   \Phi_{,\smhsp r}(t, {r})\,
 r^{D-3}\,.
 \end{align}
\textbf{Antisymmetric part.} As before, we consider first the
contribution of $\Phi_{\ah}$ : substituting $\Phi_{\ah}=-\Lambda
I_{\ah} \, \sgn(t)$ and $I_{\ah}$ from (\ref{Phi13a}), one
differentiates over $r$  using the recurrence relations for Bessel
functions,
\begin{align}\label{bessel}
\left( \frac{1}{z} \frac{\partial}{\partial z}
\right)\frac{J_{\nu}(z)}{z^{\nu}}=-\frac{J_{\nu+1}(z)}{z^{\nu+1}}\,
, \qquad\left(\frac{1}{z} \frac{\partial}{\partial z} \right)
\!\left( \vp z^{\nu}J_{\nu}(z) \right)=z^{\nu-1} J_{\nu-1}(z)\,,
\end{align}
and integrates over $k$ using \cite[eq.\,(5.11)]{GaMeS1}, to get
\begin{align}
 {\sf f}^{z}_{\ah}
=-\fr{\mu \vak^2 m }{4
 \ga}\left( \gamma^2v^2 +\fr{1}{D-2}\right)
 \left[\hsp \sgn (t) -\frac{2\ga v t }{r\sqrt{\pi}} \frac{\Gamma\left(\fr{D-1}{2}\right)}{\Gamma\left(\fr{D-2}{2}\right)}
  \; {}_2 {}F_1\!\left(  \fr12\,, \fr{D-1}{2}
\,; \fr{3}{2}\,;-\fr{\gamma^2 v^2 t^2}{r^2}\right) \right],
 \end{align}
where the limit $r \to \infty$ is to be taken. This results in
 \begin{align}\label{flow}
{\sf f}^{z}_{\ah}  = -k m\gamma\,\sgn(t)\left[(D-3){\kern
1pt}v^2+1\right]
    \,.
 \end{align}
This compensates for the extra terms in (\ref{drew3}).

\vspace{0.3cm}

\textbf{Branon part.} As was explained above, the contributions of
the branon wave to the time derivative of the wall momenta and the
lateral flux must balance each other independently, as we are going
to check now. The only nonzero are $z$ components, and  $\delta
P^z_{\bh}$ is defined in (\ref{taumn3r_1}) with the corresponding
derivative
 \begin{align} \label{fr_1b}
 F^z_{\bh}=
 2\hsp \mu  \Lambda\, \theta(t) \int  I_{\bh,\smhsp 00}(t, r)\: d^{D-2} \mathbf{r}\,.
 \end{align}
Substituting (\ref{Phi11}) and (\ref{Phi13b}) and differentiating,
one obtains
 \begin{align} \label{fr_1b_2}
 F^z_{\bh}=-
 2{\mu}\Lambda \, \Omega_{D-3}\,\theta(t) \int (kr)^{\frac{D-2}{2}} J_{\frac{D-4}{2}}(kr)\,\cos kt\, dr\,dk\,.
 \end{align}
Integration over $r$ leads to
 \begin{align} \label{fr_1b_3}
 F^z_{\bh}=-
 2{\mu}\Lambda \, \Omega_{D-3}\,\theta(t) \int \left.k^{\frac{D-4}{2}} r^{\frac{D-2}{2}} J_{\frac{D-2}{2}}(kr)\,\cos kt\,  dk
 \right|_{r=\infty}\,.
 \end{align}
On the other hand, the flux over the lateral surface $dS_r$ is
determined by the corresponding $T^{zr} $component of the brane's
stress-energy tensor and reads
 \begin{align} \label{fr_1b_4}
 \int T^{zr} dS_r= -
 {\mu} \,\Omega_{D-3}\int\left.  \Phi_{\bh,\smhsp r}(t, r)\,\delta(z)\,dz \,dt \vp\right|_{r=\infty}
 \end{align}
while the rate of its change is given by (after the trivial
$z$integration)
 \begin{align} \label{fr_1b_5}
{\sf f}^{z}_{\bh}= -
 \left.{\mu}\,  \Omega_{D-3}\,\Phi_{\bh,\smhsp r}(t,
 r)\,r^{D-3}\vp \right|_{r=\infty}\,.
 \end{align}
Differentiating it with the help of (\ref{bessel}), one arrives at
 \begin{align} \label{fr_1b_6}
{\sf f}^{z}_{\bh} =2\mu
 \Lambda \, \Omega_{D-3}\,\theta(t) \int \left.k^{\frac{D-4}{2}} r^{\frac{D-2}{2}} J_{\frac{D-2}{2}}(kr)\,\cos kt\,  dk
 \right|_{r=\infty}\,,
 \end{align}
that exactly compensates (\ref{fr_1b_3}):
 \begin{align} \label{cons_branon}
F^z_{\bh}+{\sf f}^{z}_{\bh} =0\,.
 \end{align}
 The computation of
$F^z_{\bh}$  in the closed form is presented in  Sec. VB;
%\,\ref{branon1},
 henceб
 \begin{align} \label{kozel1}
{\sf f}^{z}_{\bh}  =-  \delta \dot{P}^z_{\bh} =
  \frac{2 \hsp k m  }{\gamma }\left[ (D-2)\,\gamma^2v^2+
1\vp \right] \theta(t) \,.
 \end{align}
It is worth noting that combining two components of the lateral
force, one obtains
 \begin{align} \label{kozel2}
\frac{d}{dt} \left[\hsp  {\sf f}^{z}_{\ah} +{\sf f}^{z}_{\bh}
\vp\right]=0\,.
 \end{align}
In other words, the total  lateral $z$ force is continuous and
constant\footnote{We could apply the identity $2 \hsp\theta(t)
-\sgn(t)=1$ directly. Doing as we do, we want to emphasize that the
property is still valid at the perforation moment $t=0$.}:
 \begin{align} \label{kozel3}
  {\sf f}^{z} =  {\sf f}^{z}_{\ah} +{\sf f}^{z}_{\bh} =\left[ (D-2)\,\gamma^2v^2+
1\vp \right]  \fr{ k m  }{\gamma } \,.
 \end{align}
The same concerns the total $z$ component of momentum.

\section{Conclusions}
In this paper we have analyzed the collision problem between the
point particle and the domain wall in which no free momenta of
colliding objects can be defined, and the energy-momentum
conservation involves at any moment the contribution of the field
stresses. Generically, the stresses are nonlocal, but it turns out
that within the perturbation theory their contribution can be
unambiguously split into two parts which are effectively localized
and can be prescribed to the particle and the wall separately,
leading to the the notion of gravitational dressing. This is
somewhat similar to introduction of the potential energy in the
nonrelativistic theory, but our treatment is fully relativistic. The
dressed particle momentum involves its kinetic momentum plus its
``potential'' momentum in the field of the wall; similarly, the wall
dressed momentum involves its ``potential'' momentum in the field of
the particle. Thus our dressing is very different from the usual
dressing in the sense of adding the contribution of the proper
field. We think that such a picture may be useful also in other
situations in which the free states of the colliding objects cannot
be defined.

The second novel feature of the particle-wall collision ерфе we have
revealed here is the nonzero momentum flux through the "lateral"
surface of the world tube. Because of this flux, the divergence-free
stress tensor does not define the conserved energy-momentum charges
as the integral over timelike sections of the world tube, since the
lateral momentum flux is integrated over the time. One can still
consider the change of such integrals between the infinitesimally
closed surfaces, thus passing to the time derivatives of these
charges. Then taking into account the lateral flux we establish the
instantaneous energy-momentum balance in terms of the dressed
particle and wall momenta. Actually the nonvanishing flux arises for
the space component of the momentum orthogonal to the wall, while
the energy is still conserved in the usual sense.

The third feature, which is also fully tractable within our model,
is the excitation of the wall under the collision. Contrary to the
case of colliding particles, the wall has the internal degrees of
freedom that get excited in form of the branon wave. This excitation
consists  of two parts: one is the direct deformation of the wall in
the gravitational field of the particle, which depends on their
separation; another is the shock branon wave which starts after the
perforation and
 propagates  freely outward along the wall with the
velocity of light. The latter gives a separate contribution to the
energy-momentum which satisfies our balance equation with account
for the lateral momentum flux.

Our procedure of gravitational dressing as a relativistic
counterpart to the potential energy seems to be applicable to
collisions of particles and branes interacting via other fields. In
fact, the linearized gravity is similar to electrodynamics or any
other linear field theory. The reason for "localization" of field
stresses is that within the perturbational  treatment of collision,
the first-order field perturbations entering the field stress tensor
satisfy d'Alembert equations with localized sources. This is the
general features of classical relativistic collision problems.

\vspace{0.2cm}

\textbf{Acknowledgments.} This work was supported by the RFBR
grant 14-02-01092.

The research of PS is partially implemented under the ``ARISTEIA
II" Action of the Operational Program ``Education and Lifelong
Learning"; it is also co-funded by the European Social Fund (ESF)
and National Resources and is supported by the EU program
``Thales" ESF/NSRF 2007-2013. Finally, PS is grateful to the
non-commercial ''Dynasty'' foundation (Russian Federation) for
financial support.

\end{document}